\documentclass{ws-p8-50x6-00}

\begin{document}

\title{Future directions in astroparticle physics and the AUGER experiment}

\author{Maria Teresa Dova}

\address{Universidad Nacional de La Plata, C.C. 67, 1900 La Plata, Argentina
\\E-mail: dova@fisica.unlp.edu.ar}



\maketitle

\abstracts{The observation of cosmic ray particles with unexpected high energies is
pushing astroparticle physics into a period of rapid progress both  
theoretically and experimentally. 
Different proposed models for the generation of these particles are constrained
by the absence of the predicted GZK cutoff in the cosmic ray spectrum and by the
composition and the distribution of arrival directions observed. The database
increase due to the Pierre Auger Observatory will provide a clearer picture of
the spectral anisotropy and properties of such high energy particles, enabling
tests of their origin and nature.}

\section{The highest energy cosmic rays: Experimental results}

Since the observation of cosmic rays with energies above $10^{20} eV$ \cite{ev}, 
a considerable amount of studies have been performed in order to understand 
their origin and nature. The puzzle set by the existence of these 
ultrahigh energy cosmic rays (UHECR), which may be evidence of 
new physics or exotic particles, is nowadays one of the central 
subjects in high energy astroparticle physics. 

The energy spectrum of cosmic rays arriving to Earth extends from $10^{9} eV$ to 
$10^{20} eV$ almost continuosly over ten decades with small changes in slope in a 
power law energy spectrum: ``the knee'' appears around $10^{15.5} eV$, the second 
``knee'' at $10^{17.8} eV$ and the ``ankle''at $10^{19} eV$. Above $10^{15} eV$ 
all the measurements are indirect, the high energy particle enters in the 
atmosphere and interacts with the air molecules initiating a cascade of particles 
which can be detected by a surface array of detectors spread over a 
large area or with large aperture optical telescopes since 
during the development of the extensive air 
showers (EAS), the charged secondaries 
excite the nitrogen molecules with a subsequent emition of fluorescence light.

\begin{figure}[t]
\epsfxsize=14pc 
\epsfbox{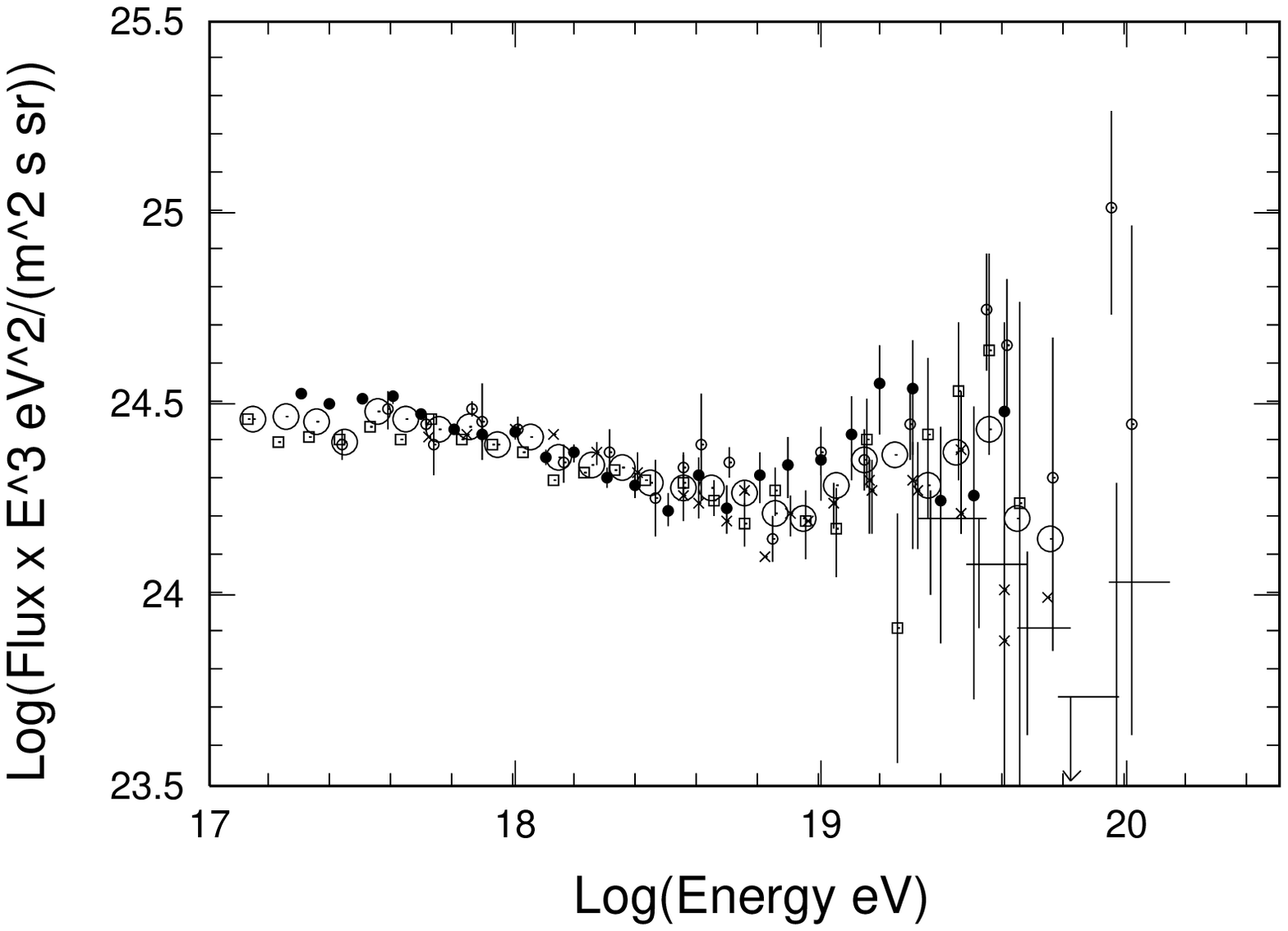} 
\epsfxsize=15pc 
\epsfbox{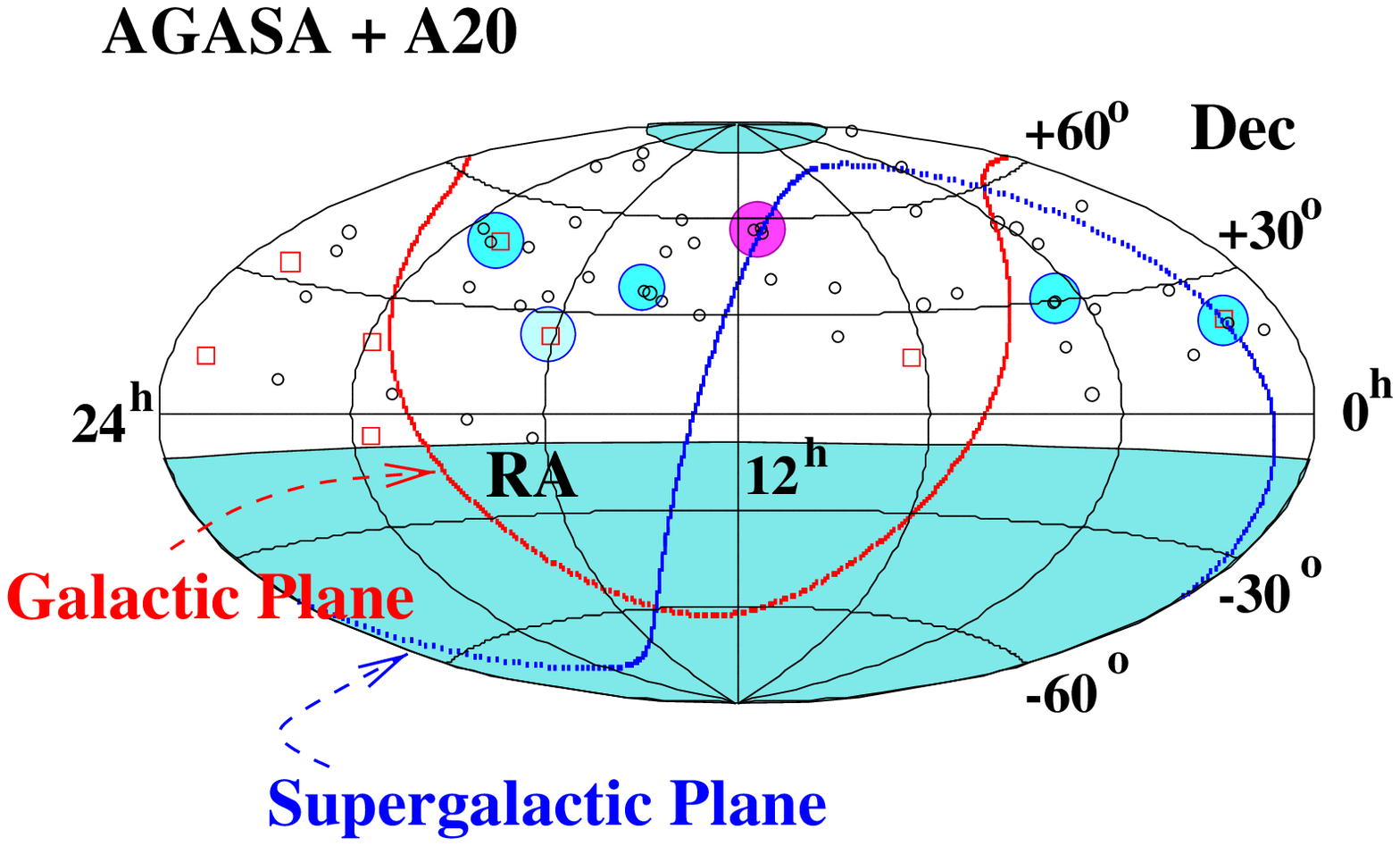}
\caption{Left:all particle
spectrum of cosmic rays for energies exceeding 100 PeV. Right:
 Arrival directions of UHECR above $10^{19}$ eV recorded by the AGASA experiment. 
\label{fig:spectrum}}
\end{figure}

Fig. 1 (left) shows the upper end of the cosmic ray spectrum where the 
differential flux is multiplied by an energy dependent 
power $E^{3}$. The compilation is from ref.\cite{xa}, with data 
from four experiments: surface arrays (Haverah Park, Yakutsk, 
AGASA) and a fluorescence detector (Fly's Eye).

The arrival directions of the events with energies above $10^{19} eV$ is 
completely consistent with an isotropic distribution with the exception 
of a few small scale anisotropies in the form of multiplets of events within the 
experimental angular resolution \cite{ta,je}. Fig.1 (right) shows the arrival 
directions of cosmic rays collected by the AGASA experiment 
\cite{ta} above $4 \times 10^{19} eV$ where four doublets and a triplet is observed. 
Shaded in grey is the area invisible to the AGASA detector. It is remarkable 
that none of those clusters is on the Galactic 
plane suggesting that UHECR are most likely extragalactic in origin. 

A crucial point in the search for the origin of UHECR is to locate their sources. The 
question is to which extent it is possible to do astronomy with the UHECR 
detected. Search for correlations of the observed multiplets with the location 
of candidate sources or with distribution of astrophysical objects in 
our neighborhood have been made with negative results. 
Besides, when doing this analysis it is very important to take into account the 
effect of the galactic and extragalactic magnetic fields \cite{to} . Magnetic deflections 
can produce additional effects: galactic magnetic fields might act as giant 
lens magnifying the CR flux coming from a single source, or even producing 
multiple images of a source.

\begin{figure}[t]
\begin{center}
\epsfxsize=15pc 
\epsfbox{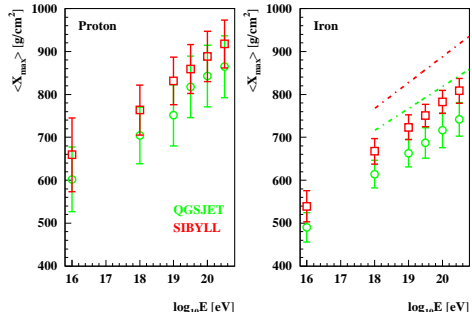} 
\end{center}
\caption{Average slant depth of maximum from simulated data. The error bars indicate the standard fluctiations.\label{fig:directions}}
\end{figure}

Another ingredient to consider in the search for the origin of these very 
energetic particles is the chemical composition of the UHECR detected. 
Experimentally, the composition can be determined, at least in a statistical manner,
 from air shower observables as 
the muon content of the shower at ground level and the position of the shower 
maximum ($X_{max}$) measured by fluorescence detectors. 
At the highest energies, the composition of the primaries seems to be consistent with 
nucleons, however the interpretation of data depends to some extent on the physics of 
the cascade included in the event generators used. Fig.2 shows simulated 
results for the average slant depth of maximum plotted vs. the logarithm of the primary 
energy for proton and iron induced showers using two well-known hadronic 
interaction models (SIBYLL and QGSJET) 
\cite{sim}.  It is evident that SIBYLL showers present higher values for 
the depth of maximum, the differences increasing with rising energy. At high energies
the primary chemical 
composition remains hidden by the hadronic interaction model.  

\section{Production and propagation of UHECR}

The existence of UHECR has motivated many detailed studies 
concerning the generation of particles with extremely high energy 
as well as their propagation in route to Earth. A complete discussion 
of most of the models for the production of UHECRs can be found in some 
recent reviews and references therein \cite{bh,ol}. Production mechanisms 
have been commonly classified into two groups: a) botton-up models, which consider 
conventional acceleration of UHECR in rapidily evolving processes in known astrophysical 
objects \cite{hi,bla,bira}. Examples are AGN radio lobes where particles can 
be accelerated via the 
first-order Fermi mechanism in the so-called hot spots, regions near neutron 
starts satisfying conditions to accelerate particles via direct electromagnetic 
acceleration; and b) top-down models suggesting that particles 
are not accelerated but rather they are stable decay products of 
supermassive particles\cite{bh}. 
Source of these exotic particles could be topological defects (TD) relics 
from early universe phase transitions associated with spontaneous symmetry 
breaking underlying unified models of high energy interactions. TD may survive to 
the present and decompose into their constituent fields. 
The supermassive particles (masses $\approx 10^{24} eV$) are supposed to decay into quarks 
which hadronize forming jets of hadrons. A general characteristic of top-down 
models is that, alongside protons, many photons and 
neutrinos are also produced given an extra signature to these processes. 
Recently, an analysis of inclined 
showers recorded by Haverah Park has been published with a new method used to 
set a limit to the photon and iron content of the UHECR \cite{ave}, setting 
important constraints to top-down models. Additional data, measurement of 
anisotropy (predicted by supermassive particles clustered as dark matter 
in the galactic halo \cite{ber}) and determination of composition is crucial 
to help solving the question of the origin of the UHECR.

\begin{figure}[t]
\epsfxsize=12pc 
\epsfbox{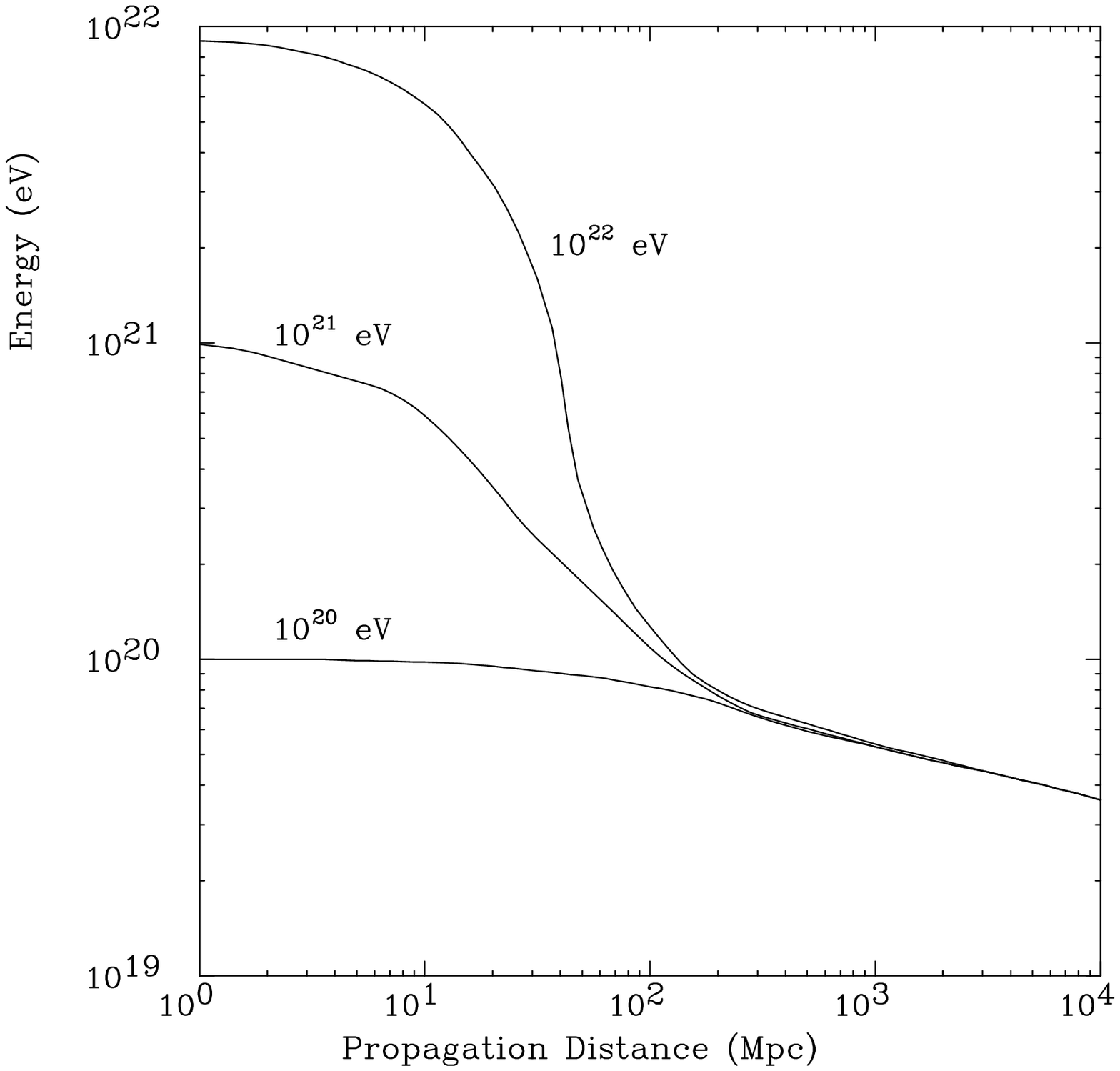} 
\epsfxsize=12pc 
\epsfbox{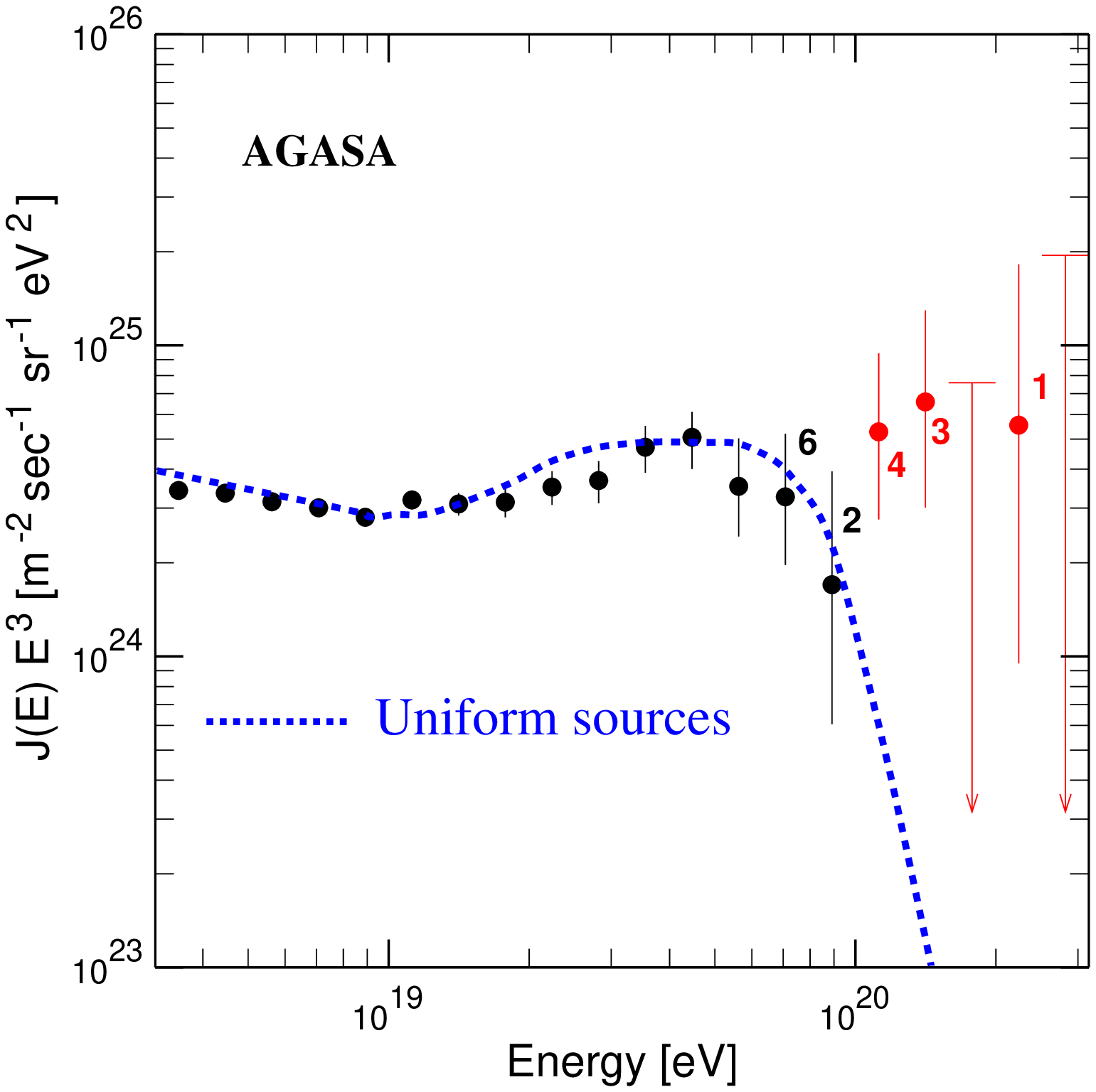} 
\caption{Left: Proton mean energy vs. propagation distance. Right: 
Cosmic ray flux spectrum from AGASA experiment shown with the shape of the universila hyphotesis spectrum.\label{fig:propa}}
\end{figure}

There is another important issue to be considered in the search for the 
origin of UHECR: the opacity of the microwave background radiation
to the propagation of UHECR. The first treatments \cite{gzk} indicated a sharp 
cutoff for cosmic rays with energies above $5 \times 10^{19}eV$ due to the 
process $\gamma + p \rightarrow \Delta \rightarrow p/n \pi $: the GZK cutoff. A similar 
phenomenon of energy degradation occurs for nuclei due to process of 
photodisintegration which is very important in the region of giant resonances.
In Figure 3(left), the energy degradation of protons in terms of their flight 
distance is shown\cite{cro}.
It can be seen that independently of the initial energy of the nucleon, 
the mean energy approach to 100 EeV after a distance of $\approx $ 100 Mpc. Since 
the energy loss mechanism depends on the nucleon energy, the emitted spectrum will 
change during the propagation. Many different calculations have been  performed 
using various techniques to study the modification of the cosmic ray 
spectrum \cite{all} and the general features are now well stablished. 
Fig. 3 (right) shows the updated AGASA measurement of the last end of 
the energy spectrum \cite{ta} together with the expected spectrum 
assuming the universal hypothesis (cosmological uniform distribution of sources) 
where the GZK cutoff is evident. Moreover significant number of events are 
observed well beyond the GZK energy. It should be mentioned that the galaxies 
found to be in the arrival directions of the multiplets are more than 70 Mpc, 
too far away to be responsible for the UHECR violating the GZK mechanisms.

\section{Pierre Auger Observatory: a hybrid detector}

The PAO has been designed  to work in a hybrid detection mode: 
particle showers are simultaneously observed by a ground array and 
fluorescence detectors \cite{pao}. The PAO is planned to measure the energy, 
arrival direction and primary species with unprecedente statistical precision. 
The observatory will be covering two sites in the Northern and Southern 
hemispheres. An engineering array 1/40th-scale, expected to be completed mid 2001,  
is under 
construction in Mendoza Province, Argentina. This site is specially 
interesting since from this part of the world, the centre of the Galaxy is visible. 

\begin{figure}[t]
\epsfxsize=10pc 
\epsfbox{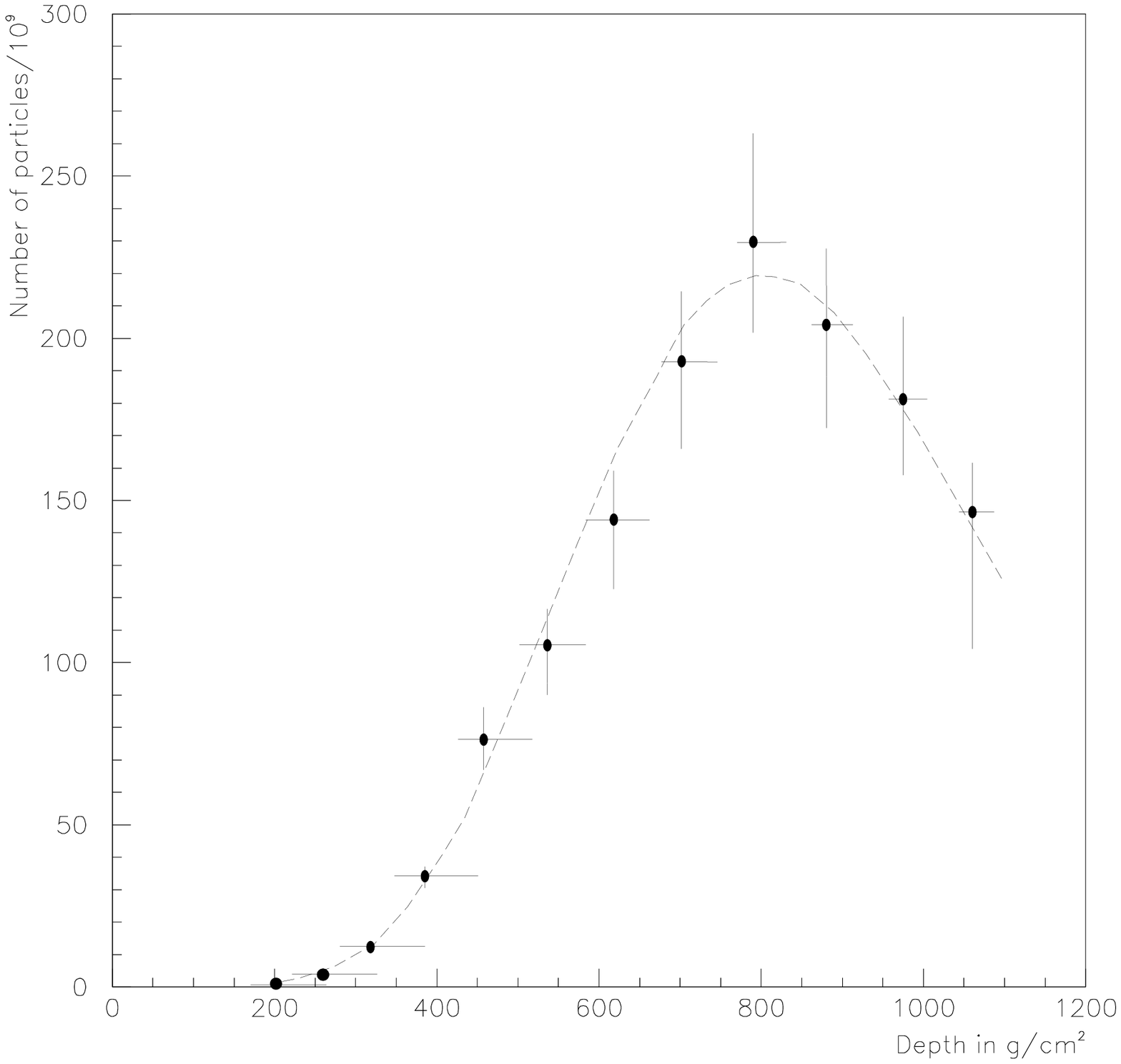} 
\epsfxsize=19pc 
\epsfbox{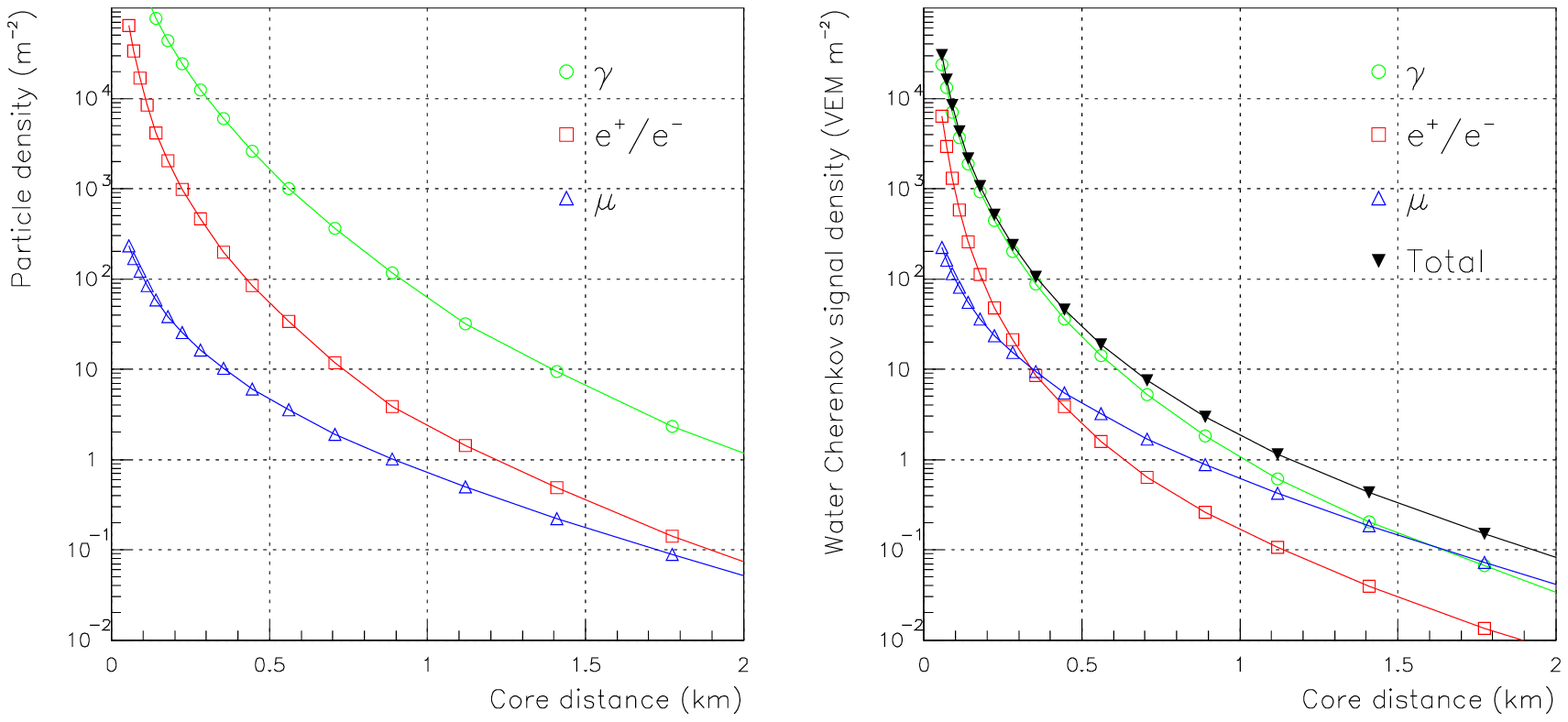} 
\caption{Left: Right: Reconstructed shower longitudinal profile for the highest energy Fly's Eye event.
Left: Particle density distributions as a function of the distance from the core.
\label{fig:longilate}}
\end{figure}

The size of the Observatory is chosen in order to collect high 
statistics above the expected GZK cutoff, with 1600 particle detectors 
covering an area of 3000 $km^2$ overviewed by four fluorescence detectors. 
Surface array stations are water Cherenkov detectors 
(a cilindrical tank of 10 $m^2$ top surface and 1.2 m height, filled 
with filtered water and lined with a highly reflective material, the 
Cherenkov light is detected by three PMTs 
installed on the top), 
spaced 1.5 km from each other in an hexagonal grid. 
These stations will operate on battery-backed solar power 
and will communicate with a central station by using wireless LAN radio links. 
Event timing will be provided through GPS receivers. 
The Observatory is completed with fluorescence detectors: three eyes will be installed at 
the periphery of the array and one at the centre. It is crucial that 
the whole array is visible by at least one of the optical detector stations.
 
The fluorescence technique is the most effective way to measure the energy 
of the primary particle. The amount of fluorescence light emitted is proportional to 
the number of charged particles in the showers allowing a direct measurement of 
the longitudinal development of the EAS in the atmosphere. Fig. 4 (left) shows the 
reconstructed longitudinal development for the Fly's Eye $ 3\times 10^{20} eV$ event. 
From this profile the position of the shower maximum $X_{max}$ can be 
obtained. The energy in the electromagnetic component is calculated by 
integrating the measured shower profile. A further correction taking into 
account the amount of unmeasured energy has to be done. PAO optical components 
will measure the EAS longitudinal profile in a similar manner. The primary energy 
can be determine by ground arrays fitting a lateral distribution function (l.d.f),
which depends on the experimental conditions, to the observed 
particles densities. Fig 4 (right) shows simulated l.d.f. of $\gamma$, electrons 
and muons at ground level for a $ 10^{19} eV$ proton shower, as well as 
the corresponding distributions convolved with the response of a typical 
PAO water Cherenkov detector. The particle density at a certain distance 
from the shower core is commonly used as an energy estimator where the conversion 
factor is evaluated from simulations. See ref. \cite{al2} 
for experimental details. 

Approximately 10\% of the showers detected by PAO 
will be observed by both surface and fluorescence  detectors allowing control of 
unwanted systematics in the primary energy determination.  The energy resolution 
in the hybrid mode will be $\approx 10\%$ and the angular resolution of 
about $0.3 ^{0}$. The hybrid  data set 
will also provide a distribution function in the multidimensional parameter space 
consisting of the quantities sensitive to the mass composition making 
possible to constrain the choice of high energy hadronic interaction models.

\section*{Acknowledgments}
I would like to thank the organizers for the kind invitation as well as 
for the financial support and the hospitality extended to me at this conference. 
A special word of thanks to Jim Cronin who made possible my participation in the 
conference. I also would like to thank A. Etchegoyen for a careful reading  
of the manuscript.

\end{document}